\documentclass{article}

\usepackage{arxiv}

\usepackage[utf8]{inputenc} 
\usepackage[T1]{fontenc}    
\usepackage{url}            
\usepackage{booktabs}       
\usepackage{amsfonts}       
\usepackage{nicefrac}       
\usepackage{microtype}      
\usepackage[noadjust]{cite}

\usepackage{graphicx}
\usepackage{bm}
\usepackage{amsfonts}
\usepackage{color}

\usepackage{amsmath}    
\usepackage{subfigure}  

\usepackage{bm}
\usepackage{amssymb}

\title{TURB-Lagr. A database of 3d Lagrangian trajectories in Homogeneous and Isotropic Turbulence}

\author{
  L. Biferale \\
  Dept. Physics and INFN\\
  University of Rome Tor Vergata and INFN, Italy\\
  \texttt{biferale@roma2.infn.it} \\
   \And
  F. Bonaccorso \\
  Dept. Physics and INFN\\
  University of Rome Tor Vergata, Italy.\\
  \texttt{fabio.bonaccorso@roma2.infn.it} \\
   \And
  M. Buzzicotti \\
  Dept. Physics and INFN\\
  University of Rome Tor Vergata, Italy.\\
  \texttt{michele.buzzicotti@roma2.infn.it} 
  \And
  C. Calascibetta \\
  Dept. Physics and INFN\\
  University of Rome Tor Vergata, Italy.\\
  \texttt{calascibetta@roma2.infn.it} 
}

\begin{document}
\maketitle

\begin{abstract}
We present TURB-Lagr, a new open database of 3d turbulent Lagrangian trajectories, obtained by Direct Numerical Simulations (DNS) of the original Navier-Stokes equations in the presence of a homogeneous and isotropic forcing. The aim is to provide the community interested in data-assimilation and/or Lagrangian-properties of turbulence a new testing-ground made of roughly 300K different Lagrangian trajectories. TURB-Lagr data are influenced by the strong non-Gaussian fluctuations characteristic of turbulence and by the rough and non differentiable fields. In addition, coming from  fully resolved numerical simulations of the original partial differential equations, they offer the possibility to apply a wide range of approaches, from  equation-free to physics-based models. TURB-Lagr data are reachable at  \url{http://smart-turb.roma2.infn.it}.

\end{abstract}


\section{Introduction}
Studying Lagrangian particle dynamics in turbulence is important for countless applications, suffice it to say that diffusion, mixing, and transport problems are all naturally addressed in the Lagrangian domain \cite{Toschi2009,Arn_odo_2008,biferale2005lagrangian,particletrapping,biferale2005multiparticle,salazar2009,lamorgese2007,yeung2006,buzzicotti2016}. Here we provide a first open database obtained from Direct Numerical Simulations (DNS) of the 3d Navier-Stokes equations with Homogeneous and Isotropic Statistics, seeded with a number of $N_p= 327680$ point-like tracers evolved up to a total time of $T = 4.5\,(\simeq 195\tau_\eta)$. Along the trajectories we provide the evolution of: particle positions, ${\mathbf X}(t)$; particles velocity and acceleration, ${\bm u}(\mathbf X(t),t)$ and $\bm a(\mathbf X(t),t)$; gradients of the underlying flow, $\partial_i u_j(\mathbf X(t),t)$.
\section{Numerical Simulations}
To generate the dataset deployed on TURB-Lagr, we have performed  a DNS of the Navier-Stokes equations (NSE) for  incompressible fluid in a triply periodic domain of size $L=2\pi$, using a standard pseudo-spectral approach fully dealiased with the two-thirds rule. The time integration has been implemented with a second-order Adams-Bashforth scheme. The simulation is done by using $N= 1024^3$ collocation points. 
A statistically steady, homogeneous and isotropic turbulent state is maintained by forcing the large scales, $0.5\leq k_{force}\leq 1.5$, of the flow via a second-order Ornstein-Uhlenbeck process  \cite{forcingsawford}.  The correlation time of the forcing is $\sim 10\tau_\eta$.  
\newline 
The NSE is:
\begin{equation}
\label{eq:nse}
\begin{cases}
\partial_t \bm{u} + \bm{u} \cdot \nabla \bm{u}  = - \nabla p  
+ \nu\Delta\bm{u} + \mathbf{F} \\
\nabla \cdot \bm{u} = 0,
\end{cases}
\end{equation}
where $p$ is the pressure and 
$\nu$ the kinematic viscosity.
The fluid density is constant and absorbed into the definition of pressure $p$. 
$\mathbf F$ is the homogeneous and isotropic forcing that drives the system to a non-equilibrium statistically steady state. We define the Reynolds number as $Re_\lambda =u_{\text{rms}} \lambda/\nu$, where $\lambda = \sqrt{5E_{tot}/\Omega} $ is the `Taylor-scale' measured from the ratio between the mean system energy and enstrophy~\cite{frisch1995turbulence}, $u_{\text{rms}}$ is the root mean square value of the velocity field. In Tab.\ref{tab:parameters} we report all parameters of the simulation.

\begin{table}[h!]
\centering
\begin{tabular}{ccccccccc}
\hline
$N$  & $L$  & $dx$  &  $dt$  & $\nu$              & $\epsilon$    & $\tau_\eta$     & $\eta$ & $Re_\lambda$ \\  
1024 & $2\pi$ & $6.1\times 10^{-3}$ & $1.5\times 10^{-4}$  & $8 \times 10^{-4}$ & $1.4 \pm 0.1$ & $0.023\pm0.003$ & $0.0042 \pm 0.0001$ & $\simeq 310$ \\ \hline
\end{tabular}
\caption{Parameters of the DNS: $N$ resolution in each dimension; $L$ physical dimension of the 3-periodic box; $dx$ grid spacing; $dt$ time step in the DNS integration; $\nu$ kinematic viscosity; $\epsilon = \nu \langle \partial_i u_j \partial_i u_j \rangle$, $\tau_\eta = \sqrt{\nu/\epsilon}$, $\eta = (\nu^3/\epsilon)^{1/4}$. In particular, we have set viscosity to have Kolmogorov scale of the order of $dx= L/N$, i.e., $\eta/dx\sim 0.7$.}
\label{tab:parameters}
\end{table}

\begin{figure}
    \centering
    \includegraphics[scale=0.49]{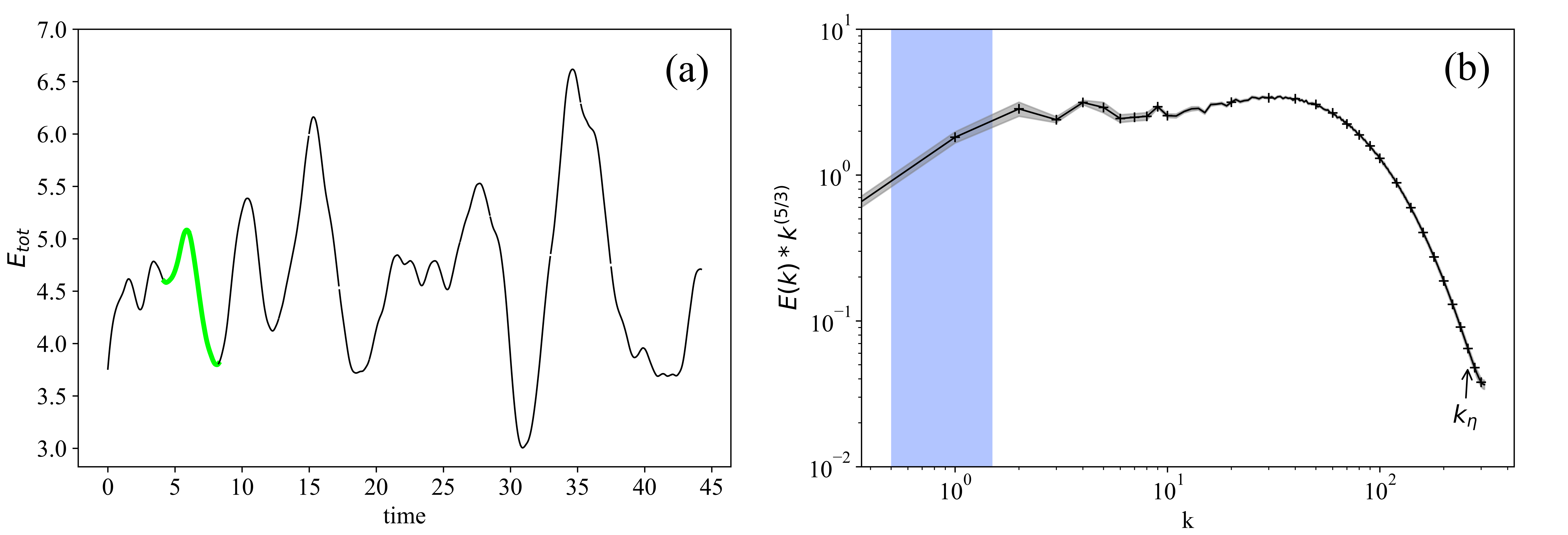}
    \caption{(Panel a) Total kinetic energy evolution, $\int dk E(k)$, for the turbulent flow generated during the simulation performed to generate the database, where $E(k)$ is the energy spectrum.
    The Lagrangian trajectories are integrated from time $t=4.2$ up to $t=8.7$ (green line).  
    (Panel b) Log-log plot of the averaged compensated energy spectrum. The blue areas indicate the forced wavenumbers, while $k_\eta \sim 230$ is the Kolmogorov dissipative wavenumber ($k_\eta=1/\eta$, see Tab\ref{tab:parameters}).} 
    \label{fig:simulation}
\end{figure}

In Fig.\ref{fig:simulation}(a) we show the total energy evolution as a function of time evolved in the full Eulerian simulation. In Fig.\ref{fig:simulation}(b) we show the energy spectrum, $E(k)=\frac{1}{2}\sum_{k\leq |\bm k | \leq k+1}{|\hat{\bm u}(\bm k)|^2}$ , compensated with the Kolmogorov K41 scaling, $k^{-5/3}$ \cite{frisch1995turbulence,pope2000turbulent}, averaged over time during the simulation, the blue shaded area identifies the frequency where the forcing is acting. 
\newline  
\hspace*{10mm}
To obtain the Lagrangian dataset, we seeded the flow with tracer particles. The particles do not react on the flow and do not interact amongst themselves. The trajectories of individual particles are described via the equation:
\begin{equation}
    \frac{d {\mathbf X}(t)}{dt}=\bm u(\mathbf X(t),t),
\end{equation}
and are integrated by using a trilinear or B-spline 6th order interpolation scheme \cite{hinsberg2012}, to obtain the fluid velocity, $\bm u(\mathbf X(t),t)$, at the particle position. Lagrangian particles evolve for roughly a large scale eddy turnover time; they start at time 4.2 and stop at time 8.7 of the Eulerian simulation just described above. In Fig.\ref{fig:simulation}(a) is highlighted in green such temporal region from which we have grounded the TURB-Lagr dataset.

\section{DataBase Description}
TURB-Lagr database is extracted from the simulation described in the previous section as follows:

\begin{itemize}
    \item During the simulation we have dumped the Lagrangian properties of $N_p=327680$ different trajectories every 15 $dt$, i.e. every $\Delta t=15dt\simeq 0.00225$ and for a total time $T=4.5$. Therefore, for each trajectory, we have stored 2000 different times; 
    \item Particles are initialized 
    with a random position extracted from the whole domain to avoid any correlation between trajectories.
    \item For each trajectory we provide particle positions, $\mathbf X(t)$, the velocity and the acceleration at the particle positions, $\bm u(\mathbf X(t),t)$, $\bm a(\mathbf X(t),t)$ and the gradients of the underlying flow, $\partial_i u_j(\mathbf X(t),t)$.   In the support folder of TURB-Lagr dataset, available in the SMART-Turb portal, we provide an example of a python file (\textit{Read\char`_Dataset.py}) to read a specific number of trajectories in the dataset. In the same file, we show how to access all the quantities dumped at each time (particle position, velocity, acceleration and gradients of the flow underlying the particle). 
    {In the dataset, particles positions are given in physical units for a box of total size $L=2\pi$.}  
\end{itemize}

The database TURB-Lagr is available for download using the SMART-Turb portal \url{http://smart-turb.roma2.infn.it}. 
\\
The dataset is provide in .h5 format. Details on how to access and read the data with a few examples can be found on the portal as support materials. 
Other data-sets concerning rotating turbulence (TURB-Rot \cite{TURB-Rot}) and Helicity data (TURB-Hel) are already available.  \\

\begin{figure}
    \centering
\includegraphics[scale=0.3]{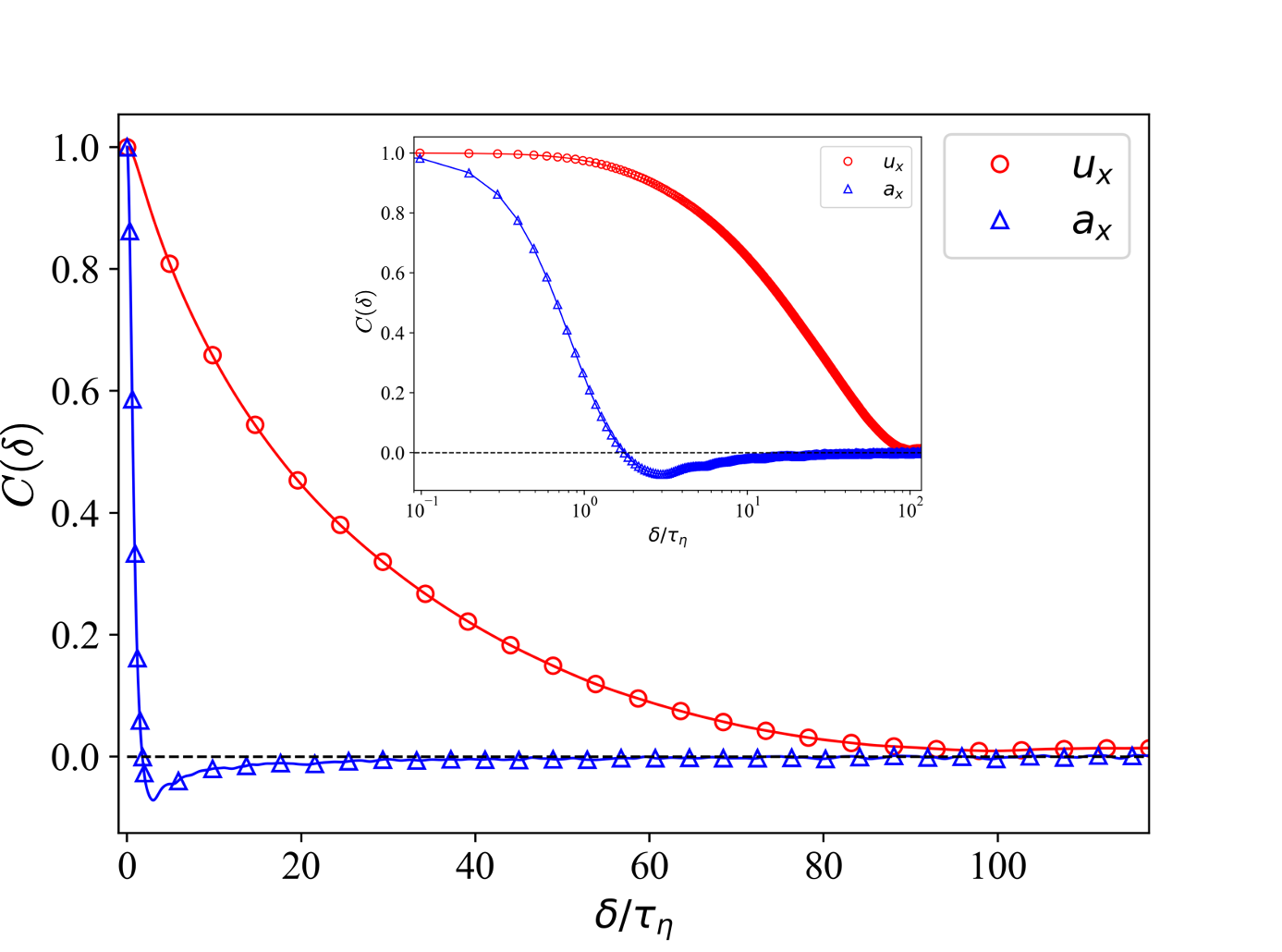}
    \caption{Correlation function of the velocity and acceleration ($x$ component). $C_s(\delta) = \big{(}\langle s(t+\delta)s(t)\rangle - \langle s\rangle^2\big{)}/\sigma^2(s)$, with $\langle...\rangle$ a mean over time, $0\leq t\leq T-\delta$, $0\leq\delta\lesssim  118 \tau_\eta$ and where $s$ is $u_x$ or $a_x$. In the inset we show the same plot in log scale.} 
\label{fig:correlation}
\end{figure}

\begin{figure}
    \centering
\includegraphics[scale=0.8]{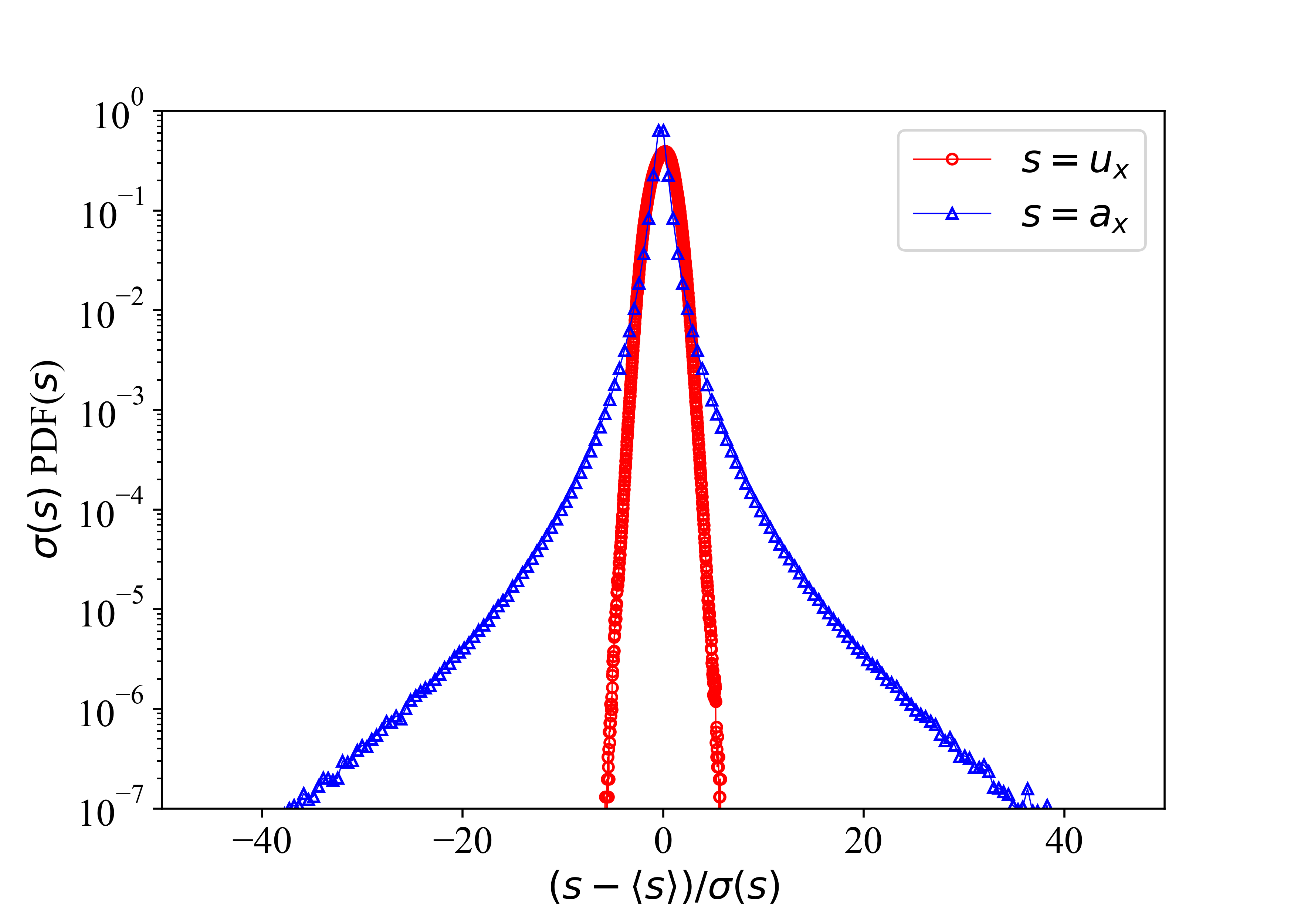}
    \caption{Standard PDF of the  velocity and acceleration, $x$-component. } 
\label{fig:pdf}
\end{figure}

\begin{figure}
    \centering
\includegraphics[scale=0.4]{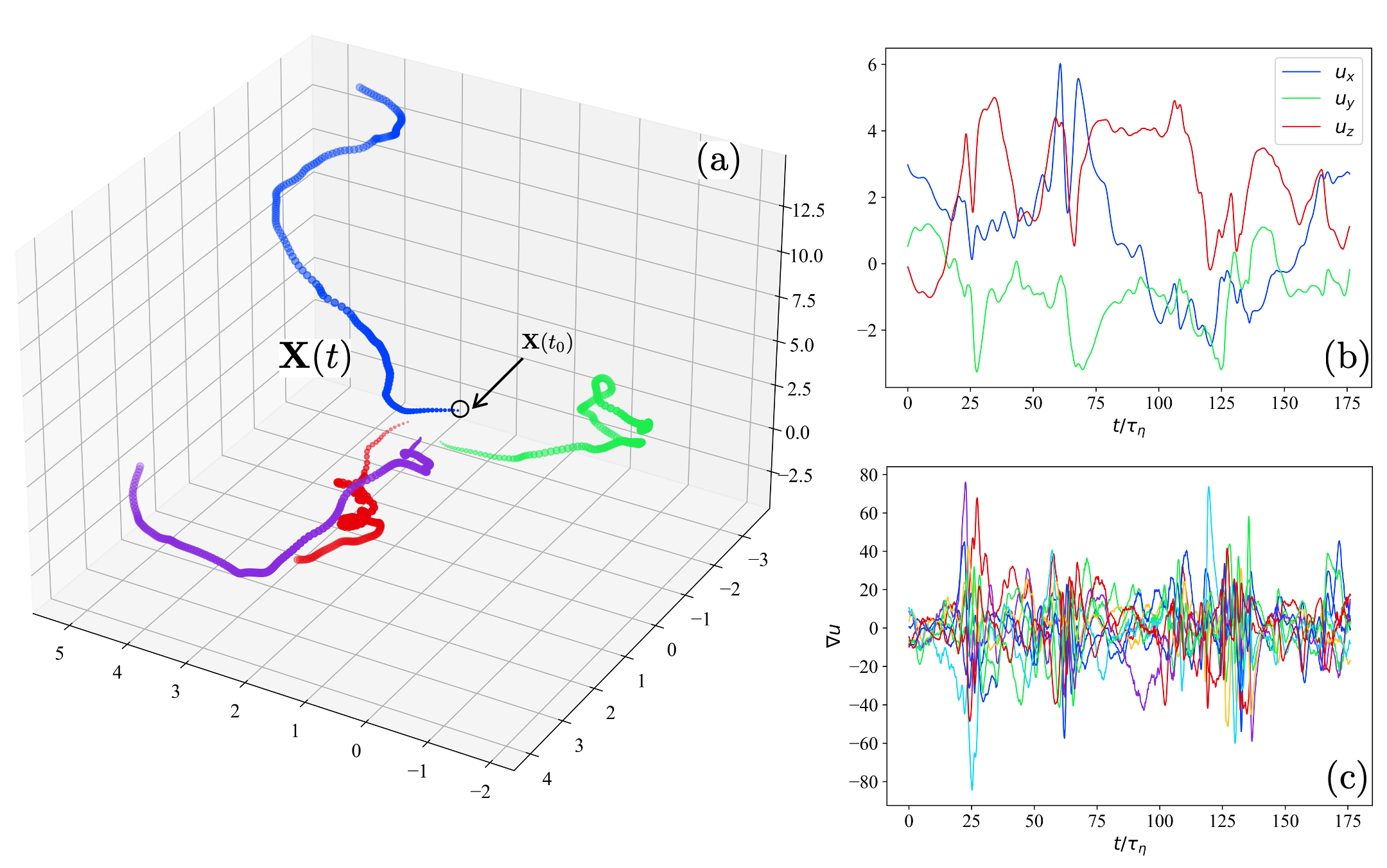}
    \caption{(Panel a) Example of evolution of 4 Lagrangian trajectories initialized with random positions. (Panel b) Evolution of the 3 components of the velocity field and (Panel c) of the 9 components of the flow gradient along the position of the blue particle in panel (a).} 
\label{fig:onetraj}
\end{figure}

\noindent Finally, In Fig.\ref{fig:correlation} we study the correlation functions of the velocity and the acceleration experienced by the particles, while in Fig.\ref{fig:pdf} we show the pdf of the same quantities. In Fig.\ref{fig:onetraj} we illustrate the evolution of velocity and gradients of the flow underlying the particle along one trajectory.

\paragraph*{Acknowledgments}
This work was supported by the European Research Council (ERC) under the European Union’s Horizon 2020 research and innovation programme (Grant Agreement No. 882340).

\bibliographystyle{unsrt}
\bibliography{references}  

\end{document}